\input harvmac
\input amssym.def
\input amssym.tex
\noblackbox
\newif\ifdraft

\catcode`\@=11
\newif\iffrontpage
\newif\ifxxx
\xxxtrue

\newif\ifad
\adtrue
\adfalse

\parindent0pt

\def\l{\lambda}

\def\a{\alpha}
\def\b{\beta}

\def\p{\partial}

\hfill

\parindent0pt

\def\{{\lbrace}
\def\}{\rbrace}

\def\R{{\Bbb R}}

\def\a{\alpha}
\def\b{\beta}

\def\l{\lambda}
\def\t{\theta}

\def\p{\partial}

\def\abstract#1{
\vskip.5in\vfil\centerline
{\bf Abstract}\penalty1000
{{\smallskip\ifx\answ\bigans\leftskip 2pc \rightskip 2pc
\else\leftskip 5pc \rightskip 5pc\fi
\noindent\abstractfont \baselineskip=12pt
{#1} \smallskip}}
\penalty-1000}

%
\lref\Moser{J.~Moser, ``On the volume elements of a manifold,'' 
Trans. Amer. Math. Soc. {\bf 120} (1965) 287.}
\lref\Lawson{H.B.~Lawson, ``Complete minimal surfaces,'' Ann. of Math. {\bf 92} (1970) 335.}
\lref\TseytlinReview{
A.~A.~Tseytlin,
``Spinning strings and AdS/CFT duality,''
arXiv:hep-th/0311139.}
%
\lref\GKP{
S.~S.~Gubser, I.~R.~Klebanov and A.~M.~Polyakov,
``A semi-classical limit of the gauge/string correspondence,''
Nucl.\ Phys.\ B {\bf 636}, 99 (2002)
[arXiv:hep-th/0204051].}
\lref\AGMOO{
O.~Aharony, S.~S.~Gubser, J.~M.~Maldacena, H.~Ooguri and Y.~Oz,
``Large N field theories, string theory and gravity,''
Phys.\ Rept.\  {\bf 323}, 183 (2000)
[arXiv:hep-th/9905111].}
%
\lref\BMN{
D.~Berenstein, J.~M.~Maldacena and H.~Nastase,
``Strings in flat space and pp waves from N = 4 super Yang Mills,''
JHEP {\bf 0204}, 013 (2002)
[arXiv:hep-th/0202021].}
%
\lref\HS{
M.~Henningson and K.~Skenderis,
``The holographic Weyl anomaly,''
JHEP {\bf 9807}, 023 (1998)
[arXiv:hep-th/9806087].}
%
\lref\Bastianelli{
F.~Bastianelli, S.~Frolov and A.~A.~Tseytlin,
``Conformal anomaly of (2,0) tensor multiplet in six dimensions and  AdS/CFT
correspondence,''
JHEP {\bf 0002}, 013 (2000)
[arXiv:hep-th/0001041];
A.~A.~Tseytlin,
``$R^4$ terms in 11 dimensions and conformal anomaly of (2,0) theory,''
Nucl.\ Phys.\ B {\bf 584}, 233 (2000)
[arXiv:hep-th/0005072].}
%
\lref\Berkovits{
N.~Berkovits,
``ICTP lectures on covariant quantization of the superstring,''
arXiv:hep-th/0209059.}
%
\lref\Metsaev{
R.~R.~Metsaev,
``Type IIB Green-Schwarz superstring in plane wave Ramond-Ramond  background,''
Nucl.\ Phys.\ B {\bf 625}, 70 (2002)
[arXiv:hep-th/0112044].}
%
\lref\AFRT{
G.~Arutyunov, S.~Frolov, J.~Russo and A.~A.~Tseytlin,
``Spinning strings in $AdS_5\times S^5$ and integrable systems,''
Nucl.\ Phys.\ B {\bf 671}, 3 (2003)
[arXiv:hep-th/0307191].}
%
\lref\ART{
G.~Arutyunov, J.~Russo and A.~A.~Tseytlin,
``Spinning strings in $AdS_5\times S^5$: New integrable system relations,''
Phys.\ Rev.\ D {\bf 69}, 086009 (2004)
[arXiv:hep-th/0311004].}
%
\lref\KS{
I.~R.~Klebanov and M.~J.~Strassler,
``Supergravity and a confining gauge theory: Duality cascades and
chiSB-resolution of naked singularities,''
JHEP {\bf 0008}, 052 (2000)
[arXiv:hep-th/0007191].}
%
\lref\MN{
J.~M.~Maldacena and C.~Nunez,
``Towards the large N limit of pure N = 1 super Yang Mills,''
Phys.\ Rev.\ Lett.\  {\bf 86}, 588 (2001)
[arXiv:hep-th/0008001].}
%
\lref\SS{
T.~Sakai and J.~Sonnenschein,
``Probing flavored mesons of confining gauge theories by supergravity,''
JHEP {\bf 0309}, 047 (2003)
[arXiv:hep-th/0305049].}
%
\lref\BEEGK{
J.~Babington, J.~Erdmenger, N.~J.~Evans, Z.~Guralnik and I.~Kirsch,
``Chiral symmetry breaking and pions in non-supersymmetric gauge /  gravity
duals,''
Phys.\ Rev.\ D {\bf 69}, 066007 (2004)
[arXiv:hep-th/0306018].}
%
\lref\KMMW{
M.~Kruczenski, D.~Mateos, R.~C.~Myers and D.~J.~Winters,
``Towards a holographic dual of large-N(c) QCD,''
arXiv:hep-th/0311270.}
%
{
\lref\HN{
J.~Hoppe and H.~Nicolai,
``Relativistic Minimal Surfaces,''
Phys.\ Lett.\ B {\bf 196}, 451 (1987).}
%
\lref\ZM{
J.~A.~Minahan and K.~Zarembo,
``The Bethe-ansatz for $N = 4$ super Yang-Mills,''
JHEP {\bf 0303}, 013 (2003)
[arXiv:hep-th/0212208].}
%
\lref\AS{
G.~Arutyunov and M.~Staudacher,
``Matching higher conserved charges for strings and spins,''
JHEP {\bf 0403}, 004 (2004)
[arXiv:hep-th/0310182].}
%
\lref\AH{
J.~Arnlind and J.~Hoppe,
``More membrane matrix model solutions, - and minimal surfaces in $S^7$,''
arXiv:hep-th/0312062.
}
%

%

\Title{\vbox{ \rightline{\vbox{\baselineskip12pt
\hbox{hep-th/0405170}}}}}
{{Spinning membranes on $AdS_p\times S^q$}
}
\vskip 0.3cm
\centerline{J.~Hoppe$^a$, S.~Theisen$^b$ } 
\vskip 0.6cm
\centerline{$^a$ \it Department of Mathematics, Royal Institute of Technology, S-10044
Stockholm, Sweden }
\vskip.2cm
\centerline{$^b$ \it Max-Planck-Institut f\"ur Gravitationsphysik,
Albert-Einstein-Institut, D-14476 Golm, Germany}

\abstract{
Minimal Surfaces in $S^3$ are shown to yield spinning membrane
solutions in $AdS_4\times S^7$.
}

\Date{\vbox{\hbox{\sl {May 2004}}
}}
\goodbreak

\parskip=4pt plus 15pt minus 1pt
\baselineskip=15pt plus 2pt minus 1pt

\noblackbox


The AdS/CFT correspondence (see \AGMOO\ for review)
offers a powerful tool to study 
interesting aspects of supersymmetric large-$N$ gauge theories beyond 
perturbation theory.  
The first stage of these developments relied mainly on 
the isomorphism between Kaluza-Klein states of classical type IIB 
supergravity compactified on $AdS_5\times S^5$ and BPS observables 
of ${\cal N}=4$ super-Yang-Mills theory in four dimensions. 
Many variations on this theme involving theories with less 
supersymmetry, with and without conformal invariance, were also 
studied, leading to quantitative results about the spectrum and phase 
structure of QCD-like theories \KS\MN\SS\BEEGK\KMMW. 

The problem to go beyond the SUGRA approximation is related to 
the difficulties to quantize string theory in 
Ramond-Ramond backgrounds. Even though a covariant quantization scheme has been 
developed \Berkovits, 
it has so far not been possible to use it to compute the 
string excitation spectrum on these backgrounds. 
An exception is the gravitational plane wave background which 
is obtained as the Penrose limit of the $AdS_5\times S^5$ vacuum of 
type IIB string theory.  
In this background light-cone quantization leads to a free theory on the 
world-sheet whose spectrum is easily computed \Metsaev. This opens the way 
to the duality between string theory and another sector of 
large-$N$ SYM, which is characterized by large $R$-charge
($\sim\sqrt{N})$ and conformal weight ($\sim\sqrt{N}$). The 
extensive activity to which this has led was initiated in \BMN.  

Studying time-dependent classical solutions of the string sigma-model 
in an $AdS_5\times S^5$ target space-time and relating them to the dual 
conformal field theory, extends the testable features of the 
duality between string theory and ${\cal N}=4$ SYM. This was proposed and 
demonstrated in \GKP. Subsequent interesting developments are summarized and 
reviewed in \TseytlinReview.  

A likely extension of these ideas is to go from strings to M-theory, where the 
fundamental objects are membranes rather than strings. In this case, 
maximally supersymmetric backgrounds, aside from eleven-dimensional 
Minkowski space, are $AdS_7\times S^4$ and $AdS_4\times S^7$.  
The former is the near-horizon limit of a stack of $N$ coincident M5 branes with 
${1\over 2}R_{AdS}=R_S=l_p(\pi N)^{1/3}$ and the latter is the near-horizon 
limit of a stack of $N$ M2 branes with $2 R_{AdS}=R_S=l_P(32\pi^2 N)^{1/6}$. 
The dualities between classical supergravity on these background and 
the conformal field theories on the world-volume of the branes which create them
has been studied. In particular for the $AdS_4\times S^7$ case, if the duality holds, 
nontrivial information about the $(0,2)$ conformal field theory of $N$ interacting 
tensor multiplets in six dimensions has been obtained, e.g. its conformal anomaly 
has been computed \HS\Bastianelli. 
Direct verifications have, however, so far been impossible, mainly 
due to the lack of knowledge of the interacting (0,2) theory.  

The problems which one 
encounters in quantizing string theory on non-trivial backgrounds are of course
much more severe in M-theory where quantization on any background is still 
elusive. The semiclassical analysis, which in the case of string theory  
provides valuable non-trivial information about the dual conformal field theory, 
can, however, be extended to M-theory. 
While the equations of motion of strings on $AdS_5\times S^5$ reduce, for 
special symmetric configurations, to classical integrable 
systems \AFRT\ART, this is 
not as simple for membranes. Also, the integrable spin-chains which 
appear in the discussion of the dual gauge theory \ZM\AS, 
have so far no known 
analogue in the (0,2) tensor theory.

In this letter we make a first step towards the semiclassical 
analysis of M-theory on $AdS_p\times S^q$ backgrounds. 
We will find that the equations of motion, upon imposing
a suitable Ansatz (analogous to the
corresponding string theory analysis, and similar to the Ansatz 
made in \HN), may be reduced
to the equations describing minimal embeddings of 2-surfaces
into higher spheres (as well as generalizations thereof).

Let us consider closed bosonic membranes in $AdS_p\times S^q$. 
Their dynamics is derived from the action 
\eqn\action{
S=\int d^3\varphi\left(\sqrt{G}+\l(\vec x^2-1)+\tilde\l(y^2-1)\right)}
where $y^\mu(\varphi^\a)\,(\mu=1,\dots,p;~\a=0,1,2)$ and 
$x_k(\varphi^\a)\,(k=1,\dots,q+1)$ are the embedding coordinates,
$\vec x^2=\sum_{k=1}^{q+1}x_k x_k,\,y^2=y^\mu y^\nu\eta_{\mu\nu}
=y_0^2+y_p^2-\sum_{\mu'=1}^{p-1}(y_{\mu'})^2$ and 
\eqn\Gab{
G_{\a\b}=\p_\a y^\mu\p_\b y^\nu\eta_{\mu\nu}-\p_\a\vec x\cdot\p_\b\vec x\,.}
The constraints 
\eqn\constraints{
y^2=1=\vec x^2}
follow by varying \action\ w.r.t. the 
Lagrange multipliers $\l$ and $\tilde\l$ while variation w.r.t. 
$y^\mu$ and $x_k$ yields the equations of motion
\eqn\eomy{
\p_\a(\sqrt{G}G^{\a\b}\p_\b y^\mu)=2\tilde\lambda y^\mu\,,}
\eqn\eomx{
\p_\a(\sqrt{G}G^{\a\b}\p_\b \vec x)=-2\lambda\vec x\,.}
Note that we take the radii of the AdS spaces and the sphere to be equal. 
It is straightforward to generalize the discussion to the case of 
unequal radii, which is the situation in the M-theory context. 
Contracting \eomy\ with $y^\mu$ and \eomx\ with $\vec x$, respectively
and using \constraints, one finds that 
\eqn\eomc{\eqalign{
2\tilde\l&=-\sqrt{G}G^{\a\b}\p_\a y^\mu\p_\b y^\nu\eta_{\mu\nu}\cr
2\l&=+\sqrt{G}G^{\a\b}\p_\a\vec x\cdot\p_\b\vec x}}
implying
\eqn\lpluslt{\eqalign{
\l+\tilde\l&=-{1\over2}\sqrt{G}G^{\a\b}\left(\p_\a y^\mu\p_\b y_\mu
-\p_\a\vec x\cdot\p_\b\vec x\right)\cr
&=-{3\over 2}\sqrt{G}\,.}}
Denoting $\varphi^0$ by $t$, let us make the Ansatz
\eqn\Ansatz{\eqalign{
&y_0=\sin(\omega_0 t),\quad y_p=\cos(\omega_0 t),\quad
y_{\mu'}=0\quad(\mu'=1,\dots,p-1)\cr
&\vec x(t,\varphi^1,\varphi^2)={\cal R}(t)\vec m(\varphi^1,\varphi^2)}}
with
\eqn\calR{
{\cal R}(t)=\pmatrix{
\cos(\omega_1 t)&-\sin(\omega_1 t)&&&\cr
\sin(\omega_1 t)&\phantom{-}\cos(\omega_1 t)&&&\cr
&&\cos(\omega_2 t)&-\sin(\omega_2 t)&\cr
&&\sin(\omega_2 t)&\phantom{-}\cos(\omega_2 t)&\cr
&&&&\ddots\cr}\,.}
Let us further demand $\dot{\vec x}\cdot \p_1\vec x=0
=\dot{\vec x}\cdot\p_2\vec x$, which, writing\hfill\break 
$\vec m^{\rm T}=\left(r_1\cos\t_1,r_1\sin\t_1,r_2\cos\t_2,r_2\sin\t_2,\dots\right)$
reads
\eqn\eomd{
\sum_{a=1}^{d\equiv[{1\over2}(q+1)]} 
\omega_a r_a^2\p_1\t_a=0=\sum_{a=1}^d\omega_a r_a^2\p_2\t_a\,.}
The world-volume metric is then block-diagonal
\eqn\wvmetric{
G_{\a\b}=\pmatrix{
\omega_0^2-\dot{\vec x}{}^2&0&0\cr
0&&\cr
\noalign{\vskip-.2cm}
&&\!\!\!\!\!\!\!\!\!-g_{rs}\cr
\noalign{\vskip-.2cm}
0&&\cr}}
with $g_{rs}=\p_r\vec x\cdot\p_s\vec x=\p_r\vec m\cdot\p_s\vec m\,\,(r,s=1,2)$ and 
$\dot{\vec x}{}^2=\sum_{a=1}^d\omega_a^2 r_a^2$. As is not difficult to see, 
\eomy\ implies that
\eqn\rhodef{
\rho:=\sqrt{G}G^{00}={\sqrt{g}\over\sqrt{\omega_0^2-\sum_{a=1}^d\omega_a^2 r_a^2}}
={g\over\sqrt{G}}}
is (a) {\it time-independent} (density). In any case,
\eqn\relomega{
\sum_{a=1}^d\omega_a^2 r_a^2+{g\over\rho^2}=\omega_0^2}
has to hold and $\tilde\l$ is determined as $-\rho\omega_0^2/2$. 

Let us now turn to the equation for $\vec x$ which determines 
$\vec m(\varphi^1,\varphi^2)$, i.e. the shape of the membrane that 
is being rotated inside $S^q$ by the orthogonal matrix 
${\cal R}(t)$ (cf.~\calR), in order to yield an extremal 
three-manifold in $AdS_p\times S^q$. With \wvmetric, \eomx\ 
becomes 
\eqn\eomxx{
{1\over\rho}\p_r\left(g {g^{rs}\over\rho}\p_s\vec x\right)=\ddot{\vec x}+{2\l\vec x\over\rho}\,.}
Due to eqs.\Ansatz,\calR\ and \eomc, implying $\ddot{\vec x}=\ddot{\cal R}(t)\vec m$, 
\eqn\eomdd{\eqalign{
{2\l\over\rho}&=\dot{\vec x}{}^2-{\sqrt{G}\over\rho}g^{rs}\p_r\vec m\cdot\p_s\vec m\cr
&=\sum_{a=1}^d\omega_a^2 r_a^2-{2g\over\rho^2}}}
\eomxx\ reduces to 
\eqn\eome{
\lbrace\lbrace m_i,m_j\rbrace,m_j\rbrace
=\left(-\omega_{(i)}^2+\sum\omega_a^2 r_a^2-{2g\over\rho^2}\right)m_i}
where 
$\omega_{(1)}=\omega_{(2)}:=\omega_1,\,\omega_{(3)}=\omega_{(4)}:=\omega_2$, etc., 
$$
g=\det(\p_r\vec x\cdot\p_s\vec x)=\det(\p_r\vec m\cdot\p_s\vec m)
=\rho^2\sum_{i<j}\lbrace m_i,m_j\rbrace^2
$$
and the (Poisson) bracket is defined as ($\epsilon^{12}=-\epsilon^{21}=1$) 
\eqn\defpb{
\lbrace f,g\rbrace={1\over\rho}\epsilon^{rs}\p_r f\p_s g}
for any two differentiable functions on the two-dimensional 
parameter manifold. The density $\rho$, though time-independent, was defined in 
\rhodef\ in a `dynamical' way, i.e. depending on $\vec x(t,\varphi^1,\varphi^2)$. 
However, due to \Moser\ we may assume it to be any given `non-dynamical' density
having the same `volume' $\int\rho(\varphi^1,\varphi^2)d^2\varphi$. This 
frees \defpb\ from its seeming $\vec x$-dependence while reducing the 
original $(\varphi^1,\varphi^2)$-diffeomorphism invariance to those preserving 
$\rho$.

Confining ourselves (for the time being) to solving \eomd\ in a trivial way by 
letting the $\t_a(\varphi^1,\varphi^2)$ be constants, i.e. 
independent of $\varphi^{1,2}$, the equations to be solved are 
\eqn\eome{
\lbrace\lbrace r_a,r_b\rbrace,r_b\rbrace=
\left(-\omega_a^2+\sum\omega_c^2 r_c^2-{2g\over\rho^2}\right)r_a\,,\qquad
a=1,\dots,d}
subject to \relomega\ and to $\sum r_a^2=1$. 
In the case of the string, rather than the membrane, this equation becomes \AFRT, for $d=3$, the 
equation of motion of the Neumann system, 
namely the constrained motion of a three-dimensional 
harmonic oscillator on the surface of a two-sphere. 

If the `spatial' frequencies $\omega_a$ are chosen to be all equal, it follows that
$\sum\omega_c^2 r_c^2=\omega^2={\rm constant}$ as well as (from \relomega)
$g/\rho^2=\omega_0^2-\omega^2={\rm const.}$ This simplifies \eome\ to 
\eqn\eomf{
\lbrace\lbrace r_a,r_b\rbrace,r_b\rbrace=-2(\omega_0^2-\omega^2)r_a}
%
which can be explicitly solved by (known) minimal embeddings of 
two-surfaces into $d=[{1\over2}(q+1)]$-dimensional unit spheres. 

To see this, one could recall \rhodef, which shows that \eomf, rewritten as
\eqn\eomg{
{1\over\rho}\p_s\left(g{g^{su}\over\rho}\p_u\vec r\right)=-2(\omega_0^2-\omega^2)\vec r\,,}
is identical to the standard `minimal surface' equation
\eqn\eomh{
{1\over\sqrt{g}}\p_s(\sqrt{g}g^{su}\p_u\vec r)=-2\vec r\,.}
This is the Euler-Lagrange equation which one obtains if one varies

\eqn\actionalt{
\int d^2\varphi\left(\sqrt{g}-\mu(\varphi)(\vec r^2-1)\right)}
w.r.t. the embedding coordinates $r_a(\varphi^1,\varphi^2)$ and the local Lagrange 
multiplier $\mu(\varphi)$ (which guarantees $\vec r^2=1$). 

Another way to show the equivalence of \eomg\ (hence \eomf) to \eomh\ is 
as follows: the results of ref.\Moser\ allow one to choose the coordinates
$\varphi^s$ in the diffeomorphism invariant equation \eomh\
such that $\sqrt{g/(\omega_0^2-\omega^2)}$ is equal to any given 
density with the same `volume' (i.e. integral over $d^2\varphi$). 
Choosing it to be $\rho$ shows that 
solutions of \eomh\ give solutions of \eomg. 
To show the converse, one notes that \eomg\ automatically 
implies that ${g\over\rho^2}=\omega_0^2-\omega^2$ 
(multiply \eomg\ by $\vec r$, and use $\vec r^2=1$ three times: 
once on the r.h.s., once for $\vec r\cdot\p_u\vec r=0$ 
and, finally, to write $\vec r\cdot \p_s\p_u\vec r$ as $-g_{su}$). 

Concerning explicit solutions of \eomf, resp. \eomh\ (from now on we put 
$\omega_0^2-\omega^2=1$ by rescaling $\rho$) let us only mention the two 
simplest ones: 
\eqn\sola{
r_1=\sin\theta\cos\varphi\,\quad
r_2=\sin\theta\sin\varphi\,,\quad
r_3=\cos\theta\,,\quad
r_{a>3}=0}
(the equator 2-sphere in $S^{d-1\geq 2},\, 
\varphi^1=\theta\in[0,\pi],\,\varphi^2=\varphi\in[0,2\pi],\,
\rho=\sin\theta$) and 
\eqn\solb{
\vec r={1\over\sqrt{2}}\left(\cos\varphi_1,\sin\varphi_1,\cos\varphi_2,\sin\varphi_2,0,\dots,0\right)}
(the Clifford-torus in $S^{d-1\geq 3}$). 
Lawson \Lawson\ proved that there exist minimal embeddings into $S^3$ 
of any topological type. 
Minimal tori in $S^7$ are given in \AH. 

\noindent
{\bf Acknowledgments:} 
We would like to thank
Joakim Arnlind, Gleb Arutyunov, Tom Ilmanen and Jan Plefka for discussions,
as well as the Albert Einstein Institute and the 
Institute for Theoretical Physics of ETH Z\"urich (J.H.)  
and the Erwin Schr\"odinger Institute (S.T.) for hospitality.

\listrefs

\bye

{\bf Bosonic membrane in $AdS_p\times S^q$}

Notation: 

Define $S^q$ and $AdS_p$ as hypersurfaces in 
$\R^{q+1}$ and $\R^{2,p-1}$, respectively: 
\eqn\hypersurface{
\eqalign{
S^q:&\qquad \vec X^2=1\cr
AdS_p:&\qquad Y^2=Y_0^2+Y_1^2-\vec Y^2=1}}
The action is 
\eqn\action{
S=\int d^3\varphi^3\left\lbrace\sqrt{G}+\lambda(\vec X^2-1)
+\tilde\lambda(Y^2-1)\right\rbrace}
$\lambda$ and $\tilde\lambda$ are Lagrange multipliers and  
\eqn\defG{
G=\det G_{\a\b}\,,\qquad
G_{\a\b}=\p_\a Y^\mu\p_\b Y^\nu\eta_{\mu\nu}-\p_a\vec X\cdot\p_\b\vec X}
is the induced metric on the world-volume. 

\listrefs

\bye